\documentstyle [12pt] {article}
\begin{document}

\def\beq{\begin{equation}}
\def\eeq{\end{equation}}
\def\Lag{{\cal L}}

\pagestyle{empty}
\rightline{\vbox{
	\halign{&#\hfil\cr
	&NUHEP-TH-92-22\cr
	&October 1992\cr}}}
\bigskip
\bigskip
\bigskip
{\Large\bf
\centerline{Effective Field Theory for Plasmas}
\centerline{at all Temperatures and Densities\footnotemark[1]}}
\footnotetext[1]{presented at the Workshop on Finite Temperature Field Theory,
	Winnipeg, July 20-25, 1992.}
\bigskip
\normalsize
\centerline{Eric Braaten}
\centerline{\sl Department of Physics and Astronomy, Northwestern University,
    Evanston, IL 60208}
\bigskip

\begin{abstract}
The solution of the plasmon problem and the subsequent development of an
effective field theory approach to ultrarelativistic plasmas are reviewed.
The effective lagrangians that summarize collective effects in
ultrarelativistic quark-gluon and electron-photon plasmas are presented.
A generalization that describes an electromagnetic plasma at all
temperatures and densities is proposed.
\end{abstract}
\bigskip

Effective field theory methods have proven to be very powerful in
treating plasmas at ultrarelativistic temperatures and densities.  This
approach was developed as a byproduct of the solution of
the long-standing plasmon problem of high temperature QCD.
It has opened the way for the systematic
calculation of many fundamental properties of ultrarelativistic plasmas
that were not feasible with previous methods.
The effective field theory approach has
recently been summarized compactly in the form of elegant effective
lagrangians.
For many applications, it would be useful to have a unified
approach that works at all temperatures and densities.  One
promising approach is to generalize the effective field theory
that describes the ultrarelativistic regime.  The effective lagrangians
for ultrarelativistic quark-gluon and electron-photon plasmas
will be reviewed below
and a generalization that describes an electromagnetic plasma
at any temperature and density will be proposed.

The plasma problem of high temperature QCD was first posed by Kalashnikov
and Klimov in 1980 and by Gross, Pisarski, and Yaffe in 1981 \cite{gpy}.
The problem was that a 1-loop calculation of the gluon damping rate,
which is proportional to the imaginary part of the gluon self energy,
gives a gauge dependent answer.
Over the next 10 years, there were about a dozen published attempts
to calculate the gluon damping rate, with almost as many different answers.
In 1989, Pisarski pointed out that a 1-loop calculation of the damping
rate is simply incomplete \cite{rdp}.
A consistent calculation to leading order
in the QCD coupling constant $g_s$ must include contributions
{}from all orders in the loop expansion.  He was able to carry out
the necessary resummation explicitly for the damping rate of a heavy quark.
The resummation consisted of replacing the gluon propagator in the 1-loop
diagram for the heavy quark self-energy by an effective gluon propagator
obtained by summing up the hard thermal
loop corrections (the terms proportional to $g_s^2 T^2$) to the gluon
self-energy.  This effective propagator was first calculated by Klimov
and by Weldon \cite{klwa}, who used it to study the propagation of gluons
and the screening of interactions in the high temperature
limit of the quark-gluon plasma.

The problem of the gluon damping rate is a little more complicated.
It is not enough to replace the gluon propagators in the 1-loop
gluon self-energy diagrams by effective propagators, because
there are also vertex corrections that are not suppressed by any powers
of $g_s$.  In particular, the 3-gluon vertex has hard
thermal loop corrections proportional to $g_s^3 T^2$ which contribute at
the same order as the bare vertex of order $g_s$.
Similarly, the 4-gluon vertex has hard
thermal loop corrections proportional to $g_s^4 T^2$, which contribute at
the same order as the bare vertex of order $g_s^4$.
A thorough diagrammatic analysis of the damping rate \cite{bpa}
by Pisarski and me revealed that the hard thermal loop corrections
to the gluon propagator, the 3-gluon vertex, and the 4-gluon vertex
are the complete set of diagrams that need to be resummed in order to
calculate the damping rate to leading order in $g_s$.
The result of this resummation was proved to be gauge invariant,
thus solving the plasmon problem.  We were also able to calculate the
damping rate explicitly \cite{bpb}, thus demonstrating that the resummation
could be used  as a practical calculational tool.
The damping rate of a quark has also been calculated \cite{kkm}
and subtleties in the proof of gauge invariance of the damping rates
have been carefully analyzed \cite{bks}.

The resummation required to solve the plasmon problem has a simple
interpretation in terms of an effective field theory.
The complete damping rate to leading order in $g_s$ is given by
the imaginary part of the 1-loop gluon self-energy diagrams, with the
gluon propagators replaced by effective propagators and with the
3-gluon and 4-gluon vertices replaced by effective vertices
obtained by adding the hard thermal loop corrections to the bare vertices.
This is equivalent to calculating 1-loop diagrams in an effective
field theory whose propagator is the effective gluon propagator of
Klimov and Weldon and whose vertices are the effective 3-gluon and 4-gluon
vertices that Pisarski and I introduced.  These propagators and vertices
are related by gauge invariance, just like their
counterparts in the QCD lagrangian.  Our analysis \cite{bpa}
revealed however that hard thermal loop corrections also appear in the
effective $n$-gluon amplitudes for $n = 5,6,7,...$. They are related
to the lower $n$-gluon amplitudes by gauge invariance, but they have no
counterparts in the QCD lagrangian.  While Pisarski and I were able to
calculate all these amplitudes explicitly, we were unable to find an
effective action which generated them.  In 1990, Taylor and Wong \cite{tw}
succeeded in writing down such an action in closed form, but it was
very cumbersome.  Finally in 1991, Frenkel and Taylor and,
independently, Pisarski and I, succeeded in writing down an
elegant effective action for the gluon amplitudes \cite{ftbpc}.

The lagrangian density which summarizes the effective field theory
for a quark-gluon plasma at ultrarelativistic temperature or density
has the form
\beq {
\Lag_{eff} \;=\; \Lag_{QCD} \;+\; \Lag_{gluon} \;+\; \Lag_{quark} \;.
} \label{LeffQCD} \eeq
The first term is the usual lagrangian density for QCD:
\beq {
{\cal L}_{QCD} \;=\; - {1 \over 2} \; {\rm tr} \; G_{\mu \nu} G^{\mu \nu}
\;+\; i \sum {\bar \psi} \gamma^\mu D_\mu \psi \;,
} \label{LQCD} \eeq
where $G_{\mu \nu} = G_{\mu \nu}^a T^a$ is the gluon field strength
contracted with generators $T^a$ that satisfy
${\rm tr}(T^a T^b) = \delta^{ab}/2$. The sum
is over $n_f$ flavors of massless quarks.
The second term in (\ref{LeffQCD}) is the thermal gluon term:
\beq {
\Lag_{gluon} \;=\; {3 \over 2} \; m_g^2
\; {\rm tr} \; G_{\mu \alpha} \left<
	{ P^\alpha P^\beta \over (P \cdot D )^2 } \right> {G^\mu}_\beta \;,
} \label{Lgluon} \eeq
where $D$ is the gauge-covariant derivative in the adjoint representation.
The angular brackets $<f(P)>$ represent the average over the spacial
directions ${\hat p}$ of the light-like 4-vector $P = (p,\vec{p})$.
The coefficient $m_g$ is the thermal gluon mass:
\beq {
m_g^2 \;=\; {g^2 \over 3} T^2 \;+\; n_f {g^2 \over 18}
	\left( T^2 + {3 \over \pi^2} \mu^2 \right) \;,
} \label{mgluon} \eeq
where $T$ is the temperature and $\mu$ is the quark chemical potential.
The quark term in (\ref{LeffQCD}) is
\beq {
\Lag_{quark} \;=\; i \; m_q^2 \; \sum {\bar \psi} \gamma_\mu
	\left< {P^\mu \over P \cdot D} \right> \psi \;,
} \label{Lquark} \eeq
where $m_q$ is the thermal quark mass:
\beq	m_q^2 \;=\; \frac{g^2}{6}
	\left( T^2 \;+\; {1 \over \pi^2} \mu^2 \right) \;.
\label{mquark}	\eeq

In astrophysical applications one encounters ultrarelativistic plasmas of
electrons and photons. The effective lagrangian that
describes such a plasma is the nonabelian version of (\ref{LeffQCD}):
\beq {
\Lag_{eff} \;=\; \Lag_{QED} \;+\; \Lag_\gamma
		\;+\; \Lag_e \;.
} \label{LeffQED} \eeq
The lagrangian for QED with massless electrons is
\beq {
\Lag_{QED} \;=\; - {1 \over 4} F_{\mu \nu} F^{\mu \nu}
\;+\; i {\bar \psi} \gamma^\mu D_\mu \psi \;.
} \label{LQED0}	\eeq
The plasma terms in the effective lagrangian (\ref{LeffQED}) are
\beq {
\Lag_\gamma \;=\; {3 \over 4} \; m_\gamma^2 \; F_{\mu \alpha}
	\left< { P^\alpha P^\beta \over (P \cdot \partial)^2 } \right>
	{F^\mu}_\beta \;,
} \label{Lgamur}	\eeq
\beq {
\Lag_e \;=\; i \; m_e^2 \; {\bar \psi} \gamma_\mu
	\left< { P^\mu \over P \cdot D} \right> \psi \;.
} \label{Lelur}	\eeq
The thermal photon mass $m_\gamma$ and the thermal electron mass $m_e$ are
\beq {
m_\gamma^2 \;=\; \frac{e^2}{9}
	\left( T^2 + {3 \over \pi^2} \mu^2 \right) \;,
} \label{mgam}	\eeq
\beq {
m_e^2 \;=\; \frac{e^2}{8}
	\left( T^2 \;+\; {1 \over \pi^2} \mu^2 \right) \;.
} \label{mel}	\eeq
The photon propagator that follows from the effective lagrangian
(\ref{LeffQED}) gives the correct dispersion relations for photons
and plasmons propagating through an ultrarelativistic plasma \cite{silin,klwa}.
The effective electron
propagator gives the correct dispersion relations for electrons
and positrons \cite{klwb},
including the effects of the thermal mass $m_e$ in (\ref{mel})
and the propagation of an extra degree of freedom
that has been christened the ``plasmino'' \cite{eb}.
An important property of these dispersion relations is that they are all
real-valued, which means that they describe the propagation of stable
quasiparticles.  Damping arises only from interactions and can be treated as
a perturbation.  The effective propagators
also correctly describe the screening of interactions
due to the plasma, not only the Debye screening of the Coulomb interaction,
but also dynamical screening of the magnetic interaction and screening
of the interaction generated by electron exchange.

The thermal photon term
(\ref{Lgamur}) in the effective lagrangian is gauge invariant because it is
constructed out of the gauge invariant field strength $F^{\mu \nu}$.
The thermal electron term (\ref{Lelur}) is also manifestly gauge invariant
and this has some important consequences.  The expansion of the covariant
derivative $D^\mu = \partial^\mu - i e A^\mu$ generates electron-photon
interactions that are related to the electron propagator by gauge
invariance.  There are not only corrections to the fundamental electron-photon
vertex, but also electron-multiphoton interactions that have no
counterpart in QED at zero temperature and density.
These effective vertices must be included for consistency in any
calculation that involves both the
effective electron propagator and the bare QED interaction.

In most applications in astrophysics, an ultrarelativistic treatment of
the plasma is insufficient.
Since an effective field theory approach seems to provide the most efficient
description of the plasma at ultrarelativistic temperatures and densities,
it would be desirable to have an effective field theory that describes the
plasma at all temperatures and densities.  The propagator of this
effective field theory should reproduce accurately
the dispersion relations for transverse photons, plasmons,
and the charged particle modes of the plasma.  The dispersion relations
should be real-valued, so that damping effects can be treated as perturbations.
The effective field theory
should also describe accurately the screening effects of the plasma.
Finally, the lagrangian for this
field theory should be gauge invariant.  An effective field theory that
satisfies all these conditions will be proposed below.

The effective lagrangian describing the propagation and interactions of
photons and electrons in the plasma will be the sum of 3 terms as in
(\ref{LeffQED}).  The QED term is the same as (\ref{LQED0}), except that the
bare electron mass can no longer be neglected:
\beq {
\Lag_{QED} \;=\; - {1 \over 4} F_{\mu \nu} F^{\mu \nu}
\;+\; i {\bar \psi} \gamma^\mu D_\mu \psi
\;-\; m_e {\bar \psi} \psi \;.
} \label{LQED}	\eeq
Here and below, $m_e$ represents the bare electron mass, $m_e = 0.511$ MeV,
and not the thermal mass given in (\ref{mel}).
For a plasma that contains charged particles that are not
ultrarelativistic, the photon term (\ref{Lgamur}) must be generalized to
\beq { \Lag_\gamma \;=\;
	e^2 \; F_{\mu \alpha} \left( \sum_i Q_i^2
	\int {d^3p \over (2 \pi)^3} {1 \over 2 E} f_i(E)
	{ P^\alpha P^\beta \over (P \cdot \partial)^2 }
	\right) {F^\mu}_\beta \;,
} \label{Lgam}	\eeq
where the sum is over all the spin states of the charged particles
and $Q_i$ is the charge of the $i$'th type of particle in units of the
proton charge $e$.  The integral is over the phase space of
$P = (E, \vec{p})$, where $E = \sqrt{m_i^2 + p^2}$ and $m_i$ is the
mass of the $i$'th particle.  The thermal distribution function $f_i(E)$
is the Bose distribution if the particle is a boson and the Fermi
distribution  if it is a fermion.
In the case where the only charged particles in the plasma
are ultrarelativistic electrons and positrons, each having two spin states,
the effective lagrangian (\ref{Lgam}) reduces to (\ref{Lgamur}).

The effective lagrangian (\ref{Lgam}) takes into account to leading order
in the electromagnetic coupling $e^2$ the forward scattering amplitudes
of pointlike charged particles in the plasma.
For example, an electron in the plasma  with 4-momentum $P$ can interact
with the electromagnetic field in such a way that it is scattered back into
the same momentum state $P$.  Representing the electromagnetic field
by an off-shell  photon with 4-momentum $K$, this process is
\beq {
\gamma^*(K) \; e^-(P) \; \rightarrow \; \gamma^*(K) \; e^-(P) \;.
} \label{fsel} \eeq
Since the initial and final states both consist of a virtual photon
in a plasma at equilibrium, these forward scattering amplitudes should be added
coherently over the thermal distribution of the momentum $P$
and over all the types of charged particles as well.  The mathematical
approximation that leads to the effective lagrangian (\ref{Lgam})
is to take $K^2 << P \cdot K$ in the forward scattering amplitudes.
This is equivalent to calculating the forward scattering amplitudes
at zero temperature and electron density \cite{gb}.
For the values of the photon momentum $K$
for which plasma corrections to the photon
propagator are important, the condition $K^2 << P \cdot K$
is well-satisfied for almost all the momenta $P$
allowed by the thermal distribution $f_i(E)$ in (\ref{Lgam}).

There are 3 limiting cases of an
electromagnetic plasma that have been studied thoroughly:  the
nonrelativistic limit, the degenerate limit, and the ultrarelativistic limit.
One can verify that the term (\ref{Lgam}) accurately describes
the propagation of photons and plasmons and the screening of the
electromagnetic field in each of these 3 limits.  As illustrations,
I consider the plasma frequency $\omega_{pl}$ and the Debye screening mass
$k_D$.  The general formulas for these quantities that follow from the
effective lagrangian (\ref{Lgam}) are
\beq {
\omega_{pl}^2 \;=\; {2 \alpha \over \pi} \sum_i Q_i^2
\int_0^\infty dp \; {p^2 \over E} f_i(E)
\left( 1 - {1 \over 3} v^2 \right)  \;,
} \label{wpl} \eeq
\beq {
k_D^2 \;=\; {2 \alpha \over \pi} \sum_i Q_i^2
\int_0^\infty dp \; {p^2 \over E} f_i(E)
\left( 1 + {1 \over v^2} \right)  \;,
} \label{kD} \eeq
where $v = p/E$ is the velocity of the charged particle.
For an ultrarelativistic plasma of electrons and positrons,
(\ref{wpl}) and (\ref{kD}) reduce to
$\omega_{pl} \rightarrow m_\gamma$ and $k_D \rightarrow \sqrt{3} m_\gamma$,
where $m_\gamma$ is the thermal photon mass given in (\ref{mgam}).
In the nonrelativistic limit, the expressions (\ref{wpl}) and (\ref{kD})
reduce to the classic results:
\beq {
\omega_{pl}^2 \;\rightarrow\; \sum_i Q_i^2 \; {4 \pi \alpha n_i \over m_i} \;,
} \label{wplcl} \eeq
\beq {
k_D^2 \;\rightarrow\; \sum_i Q_i^2 \; {4 \pi \alpha n_i \over T} \;,
} \label{kDcl} \eeq
where $n_i$ is the number density of the appropriate spin state $i$.
In the degenerate limit, they also reduce to the correct expressions:
\beq {
\omega_{pl}^2 \;\rightarrow\; \sum_i Q_i^2 \; {2 \alpha \over 3 \pi}
		p_{Fi}^2 v_{Fi} \;,
} \label{wpld} \eeq
\beq {
k_D^2 \;\rightarrow\; \sum_i Q_i^2 \; {2 \alpha \over \pi}
	{p_{Fi}^2 \over v_{Fi}} \;,
} \label{kDd} \eeq
where $p_{Fi} = (6 \pi^2 n_i)^{1/3}$ is the Fermi momentum of
the spin state $i$ and $v_{Fi}$ is the corresponding Fermi velocity.
Since the effective lagrangian correctly describes the 3 limiting cases
and interpolates smoothly between them, it should be accurate
at all temperatures and densities.

An electron term in the effective lagrangian can be derived by the same
methods as the photon term. It must take into account forward scattering
amplitudes involving photons, electrons, and positrons in the plasma.
Representing the electron
field by an off-shell electron of momentum $K$, the relevant
processes are
\beq {
e^*(K) \; \gamma(Q) \; \rightarrow \; e^*(K) \; \gamma(Q) \;,
} \label{fsgam} \eeq
\beq {
e^*(K) \; e^\pm(P) \; \rightarrow \; e^*(K) \; e^\pm(P) \;.
} \label{fsee} \eeq
Calculating the forward scattering amplitudes
at zero temperature and electron density as in Ref. \cite{gb},
they can be taken into account in the effective lagrangian
by adding the following electron term:
\beq {
{\cal L}_e \;=\;  i \; e^2 \; {\bar \psi} \left(
2 \int {d^3q \over (2 \pi)^3} {1 \over 2 q} f_\gamma(q)
		{Q \cdot \gamma \over Q \cdot D}
\;+\; \sum_\pm \int {d^3p \over (2 \pi)^3} {1 \over 2 E} f_{e^\pm}(E)
		{P \cdot \gamma \mp 2 m_e \over P \cdot D \pm i m_e^2}
		\right) \psi \;,
} \label{Lel} \eeq
where $Q = (q, {\vec q})$ and $P = (E, {\vec p})$ with $E = \sqrt{p^2 +
m_e^2}$.
The resulting effective propagator describes not only the propagation of
electrons and positrons, but the plasmino as well.
Effective electron-photon and electron-multiphoton
vertices appear upon expanding the covariant derivative
$D^\mu = \partial^\mu - i e A^\mu$ in
(\ref{Lel}).  In any calculation that involves both the effective
electron propagator and the bare electron-photon vertex,
gauge invariance requires that
the effective vertices also be included for consistency.
The effective lagrangian (\ref{Lel})
reduces to (\ref{Lelur}) in the limit of ultrarelativistic temperature
or electron density.  Its behavior in the nonrelativistic and degenerate
limits has not yet been studied.

The effective lagrangian consisting of (\ref{LQED}), (\ref{Lgam}), and
(\ref{Lel}) should provide a useful starting point for describing an
electromagnetic plasma at all temperatures and densities.  It sums up
the coherent effects of forward scattering amplitudes into effective
propagators and vertices.  The effective propagators incorporate the
the screening of the electromagnetic interaction by the plasma,
as well as its effects on the
dispersion relations for the propagating modes.
The effective vertices in the electron term
are required by gauge invariance.  The effective
field theory approach has proved to be very powerful for calculating the
properties of the plasma in the ultrarelativistic limit.
The effective field theory presented above allows a smooth interpolation
between these calculations and the corresponding results in
the nonrelativistic and degenerate limits.
Thus it allows for a unified description of an electromagnetic plasma
at all temperatures and densities.

\vfill\eject


\begin{thebibliography}{99}
%
\bibitem{gpy}
{O.K. Kalishnikov and V.V. Klimov, {\it Sov. J. Nucl. Phys.}
		{\bf 31}, 699 (1980);
	D.J. Gross, R.D. Pisarski, and L.G. Yaffe,
		{\it Rev. Mod. Phys.} {\bf 53}, 43 (1981).}
%
\bibitem{rdp}
{R.D. Pisarski, {\it Phys. Rev. Lett.} {\bf 63}, 1129 (1989).}
%
\bibitem{klwa}
{V. V. Klimov, {\it Sov. Phys. J.E.T.P.} {\bf 55}, 199 (1982);
	H. A. Weldon, {\it Phys. Rev.} {\bf D26}, 1394 (1982).}
%
\bibitem{bpa}
{E. Braaten and R.D. Pisarski, {\it Phys. Rev. Lett.} {\bf 64}, 1338 (1990);
		{\it Nucl. Phys.} {\bf B337}, 569 (1990);
		{\it Nucl. Phys.} {\bf B339}, 310 (1990).}
%
\bibitem{bpb}
{E. Braaten and R.D. Pisarski, {\it Phys. Rev.} {\bf D42}, 2156 (1990).}
%
\bibitem{kkm}
{R. Kobes, G. Kunstatter, and K. Mak, {\it Phys. Rev.} {\bf D45}, 4632 (1992);
E. Braaten and R.D. Pisarski, {\it Phys. Rev.} {\bf D46}, 1829 (1992).}
%
\bibitem{bks}
{R. Baier, G. Kunstatter, and D. Schiff,
{\it Phys. Rev.} {\bf D45}, R4381 (1992);
A. Rebhan, CERN preprint CERN-TH 6434/92 (January 1992);
H. Nakkagawa, A. Niegawa, and B. Pire, Ecole Polytechnique preprint
	CPTH-0292 (February 1992).}
%
\bibitem{tw}
{J.C. Taylor and S.M.H. Wong, {\it Nucl. Phys.} {\bf B334}, 199 (1991).}
%
\bibitem{ftbpc}
{J. Frenkel and J. C. Taylor, {\it Nucl. Phys.} {\bf B374}, 156 (1992);
	E. Braaten and R.D. Pisarski, {\it Phys. Rev.} {\bf D45}, R1827 (1992).}
%
\bibitem{silin}
{V.P. Silin, {\it Sov. Phys. J.E.T.P.} {\bf 11}, 1136 (1960).}
%
\bibitem{klwb}
{V.V. Klimov, {\it Sov. J. Nucl. Phys.} {\bf 33}, 934 (1981);
	H.A. Weldon, {\it Phys.  Rev.} {\bf D26}, 2789 (1982).}
%
\bibitem{eb}
{E. Braaten, {\it Astrophys. J.} {\bf 392}, 70 (1992);
	G. Baym, J.-P. Blaizot, and B. Svetitsky,
		Saclay preprint S.Ph-T/92-027 (May, 1992).}
%
\bibitem{gb}
{G. Barton, {\it Ann. Phys. (N.Y.)} {\bf 200}, 271 (1990).}
%
\end{thebibliography}
\end{document}